\documentclass[3p]{elsarticle}

\usepackage[dvipsnames]{xcolor}
\usepackage[most]{tcolorbox}

\usepackage{lineno,hyperref}
\usepackage{siunitx}
\usepackage{float}
\usepackage{color}
\usepackage{colortbl}
\usepackage{amsmath}
\usepackage{amssymb}
\usepackage{booktabs}
\biboptions {sort&compress}

\usepackage[symbol]{footmisc}

\modulolinenumbers[5]

\journal{Journal of Structural Control and Health Monitoring}

\begin{document}

\begin{frontmatter}

\begin{tcolorbox}[colback=white,
	colframe=red,
	width=16cm,
	]
	This is the pre-peer reviewed version of the following article:\\ "Schleiter S., Altay O. Identification of abrupt stiffness changes of structures with tuned mass dampers under sudden events. \textit{Struct Control Health Monit.}. 2020",\\ which has been published in final form at \url{https://onlinelibrary.wiley.com/doi/full/10.1002/stc.2530}. This open access article may be used for non-commercial purposes in accordance with Wiley Terms and Conditions for Use of Self-Archived Versions.

\end{tcolorbox}

\title{Identification of Abrupt Stiffness Changes of Structures with Tuned Mass Dampers under Sudden Events}

\author{Simon Schleiter}
\author{Okyay Altay\corref{mycor}}
\cortext[mycor]{Corresponding author}
\ead{altay@lbb.rwth-aachen.de}
\address{Department of Civil Engineering, RWTH Aachen University\\ Mies-van-der-Rohe-Str. 1, 52074 Aachen, Germany}

\begin{abstract}
This paper presents a recursive system identification method for multi-degree-of-freedom (MDoF) structures with tuned mass dampers (TMDs) considering abrupt stiffness changes in case of sudden events, such as earthquakes. Due to supplementary non-classical damping of the TMDs, the system identification of MDoF+TMD systems disposes a challenge, in particular, in case of sudden events. This identification methods may be helpful for structural health monitoring of MDoF structures controlled by TMDs. A new adaptation formulation of the unscented Kalman filter allows the identification method to track abrupt stiffness changes.
The paper, firstly, describes the theoretical background of the proposed system identification method and afterwards presents three parametric studies regarding the performance of the method. The first study shows the augmented state identification by the presented system identification method applied on a MDoF+TMD system. In this study, the abrupt stiffness changes of the system are successfully detected and localized under earthquake, impulse and white noise excitations. The second study investigates the effects of the state covariance and its relevance for the system identification of MDoF+TMD systems. The results of this study show the necessity of an adaptive definition of the state covariance as applied in the proposed method. The third study investigates the effects of modeling on the performance of the identification method. Mathematical models with discretization of different orders of convergence and system noise levels are studied. The results show that, in particular, MDoF+TMD systems require higher order mathematical models for an accurate identification of abrupt changes.
\end{abstract}


\end{frontmatter}



\section{Introduction}\label{sec1}

In the past decades, the importance of system identification in civil engineering has grown continously. Response measurements using accelerations, velocities, etc. are common and can be applied for the identification of important parameters, such as the natural frequencies and damping ratios of linear structures. However, the identification of nonlinear structures, including system damages, got more attention recently. As Devin and Fanning\cite{Devin.2019} mentioned, even partition walls or in general non-structural elements have high influence on natural frequencies and damping ratios, so that, for instance in case of an earthquake, even small damages can deteriorate the dynamic performance. The damage detection is, therefore, an important research field of nonlinear system identification.

Numerous system identification methods have been proposed so far, which can be split into offline and online, i.e. recursive, methods. For offline methods, a whole data set of system responses is required, while recursive methods enable stepwise system identification based on \textit{a priori} system informations and states. One field of offline identification is the operational modal analysis (OMA)\cite{Brincker.2015}, which can be further divided in time domain methods, including autoregressive moving average (ARMA) methods\cite{Soderstrom.1994} or stochastic subspace identification (SSI)\cite{Brincker.2015}, and the frequency domain methods, such as the frequency domain decomposition (FDD)\cite{Brincker.2001,Schleiter.2018}, respectively. The aim of the OMA is the \textit{a posteriori} system identification under white noise input signals and, therefore, output-only measurements.

On the other hand, online identification methods are necessary to identify nonlinear system changes in real-time. In contrast to offline identification methods, only time domain methods can be applied for the online identification, since basic frequency domain methods are generally nonrecursive. Some examples for the time domain methods, which were successfully applied to civil engineering structures, are, for instance, least square estimation (LSE)\cite{Smyth.1999}, particle filters (PF)\cite{Chatzi.2009} and Kalman filter (KF)\cite{Kalman.1960}. Among these methods, Kalman filter methods, in particular, have become one of the common methods for system identification, since a combined parameter and state estimation is possible and the computational cost is moderate.

The KF is a recursive method for estimating states, considering e.g. displacements and velocities, and can be applied for any type of excitation signal. However, only linear system behavior can be covered by the KF.
On this account, the extended Kalman filter (EKF) has been proposed\cite{Jazwinski.1997,Haykin.2001}. In general, the EKF uses the same concept as the KF, except for the nonlinear state and observation equations, which are linearized in each calculation step by setting up their Jacobians. Although nonlinear systems can be covered by this linearization, EKF has two main disadvantages: Firstly, it is very costly to set up the Jacobians in each time step, which makes a recursive application more difficult, and secondly for highly nonlinear systems the linearization approach is not accurate enough. Therefore, a further development, the unscented Kalman filter (UKF) by Julier \textit{et al.}\cite{Julier.1997,Julier.2004}, has been proposed, avoiding the disadvantages of the EKF. Instead of the linearization, an unscented transformation (UT), based on sampling points is used and thus systems with higher nonlinearities can be covered.

In addition, the main advantage of both EKF and UKF is the possibility of a joint state and parameter estimation\cite{Wan.14Oct.2000,Hoshiya.1984}, which includes, besides the common displacement and velocity states, also system parameters, such as stiffness coefficients. The joint state and parameter estimation of the UKF was applied to multi-degree-of-freedom (MDoF) civil engineering structures both numerically\cite{Miah.2015,Roffel.2014} and experimentally\cite{Miah.2017}. Within these studies, however, only nonlinearities due to initial stiffness deviations were investigated and apart from that, systems were assumed to behave linearly. Further studies have shown the applicability of the parameter estimation using UKF to the nonlinear Bouc-Wen material behavior\cite{Wu.2007}, as well as for negative stiffness devices in frame structures\cite{Erazo.2018}.

To cover abrupt system changes of MDoF structures, the UKF or respectively EKF has to be adaptive. On this account several attempts have been made in the past. For instance, Yang \textit{et al.}\cite{Yang.2006} propose a recursively determined forgetting factor introduced in the EKF, which is calculated by an optimization step based on stiffness estimates. Lei \textit{et al.}\cite{Lei.2016}, instead, propose a three step algorithm, where, firstly, the initial system parameters are identified by an EKF. For each following time steps, the damages are detected by the innovation error, subsequently, identified and localized by an optimization step and, finally, the new states are identified using a KF. In contrast, Bisht and Singh\cite{Bisht.2014} propose an adaptive unscented Kalman filter (A-UKF). Damages herein are detected by an adaptation criterion based on the innovation error and \textit{a posteriori} known system response data. Finally, the estimation of abrupt changes is enabled by the adaptation of the state covariance. Although Rahimi \textit{et al.}\cite{Rahimi.2017} use a similar approach like Bisht and Singh, they modified the adaptation criterion, where the adaptation threshold is calculated non-recursively based on sensitive floating variances.

Tuned mass dampers (TMDs) introduce supplementary damping and restoring forces on structures. Therefore, MDoF structures with TMDs respond to dynamic excitations with lower amplitudes and shorter vibration duration than systems without TMDs. Consequently, in particular in case of sudden events with abrupt changes, the system identification performance of MDoF structures with TMDs is expected to deteriorate. On this account, the accuracy of stiffness identification is more challenging for systems with TMDs in contrast to those without additional damping devices. However, to the best of authors' knowledge, no previous study has investigated the performance of the recursive system identification approaches for the estimation of abrupt changes of MDoF+TMD systems. In this context, in particular, the accuracy of the chosen mathematical model is important. The previous studies mostly used linearized mathematical models, which cannot reach the required accuracy level for MDoF+TMD systems. A comparison and a careful choice of existing mathematical models in nonlinear system identification is absolutely necessary. Furthermore, the so far proposed adaptation algorithms for the UKF either require a completed system response time-window in a nonrecursive manner or include highly sensitive nonrobust calculation procedures. Since for real-time measurement scenarios the response data is available only stepwise and signals are biased by noise, a robust recursive adaptation criterion is required.

This paper presents an UKF-based system identification method for MDoF+""TMD systems. In Section \ref{sec2}, an adaptive approach with robust, recursive algorithms is proposed, which enables the detection of abrupt stiffness changes of MDoF+TMD systems during sudden events with a high-level accuracy. In particular, the proposed approach needs no special knowledge of \textit{a posteriori} system responses. Thus, a constant adaptation criterion based on known sensor properties is driven by statistical signal properties. In Section \ref{sec3}, the presented method is investigated on a MDoF+TMD system by three parametric studies. In the first study, using the proposed system identification method a stiffness and state estimation considering abrupt stiffness changes is performed for several load scenarios. The remaining two studies focus on the filter setup of the system identification method and its influence regarding the identification performance. Therefore, the state covariance influence regarding the identification speed, especially in terms of TMD equipped structures, is investigated in the second study. Finally, the accuracy of four Taylor expansion based mathematical models of different orders of convergences are analyzed in the third study. In particular, the relationship between the system noise level and mathematical model is explored. A conclusion of the work is presented in Section \ref{sec4}.


\section{System identification method for MDoF structures with TMDs}\label{sec2}
%
\subsection{System identification method}\label{sec2.1}

For the system identification of MDoF+TMD systems, an unscented Kalman filter (UKF)\cite{Julier.1997} based system identification method is proposed. Similar to the linear KF, the UKF consists of a prediction as well as a correction step. To cover the nonlinearities the UT is applied. In the UT, sampling points $\mathbf{\tilde{X}}_k^i$ of size $n$, where $\tilde{\cdot}$ denotes corrected values, are created on the basis of the known mean $\mathbf{\tilde{x}}_k$ and the current state covariance $\mathbf{\tilde{P}}_k$, which is assumed to be Gaussian distributed. For the calculation of the sampling points, weighting factors $W_m^i$ and $W_c^i$ are introduced for mean and respectively covariance values. The scaling parameters $\lambda$, $\alpha$, $\beta$ and $\kappa$ are standard values and are mostly chosen based on a Gaussian distribution\cite{Wan.14Oct.2000}:
\begin{equation}
\mathbf{\tilde{X}}_k^i=\mathbf{\tilde{x}}_k\pm\left(\sqrt{(n+\lambda)\mathbf{\tilde{P}}_k}\right)_i\hspace{7mm}i=1,...,2n\\
\end{equation}
\begin{equation}
\mathbf{\tilde{X}}_k^0=\mathbf{\tilde{x}}_k\\
\end{equation}
\begin{equation}
W_m^0=\frac{\lambda}{n+\lambda}\\
\end{equation}
\begin{equation}
W_c^0=\frac{\lambda}{n+\lambda}+1-\alpha^2+\beta\\
\end{equation}
\begin{equation}
W_m^i=W_c^i=\frac{1}{2(n+\lambda)}\\
\end{equation}
\begin{equation}
\lambda=\alpha^2(n+\kappa)-n\\
\end{equation}

Using a nonlinear time variant state equation $f(\cdot)$ all sampling points are transformed to the estimates $\mathbf{\hat{X}}_{k+1}^i$ at the next time step $k+1$, where $\hat{\cdot}$ denotes estimation values. Under assumption of a Gaussian distribution for both time steps $k$ and $k+1$ the state estimate $\mathbf{\hat{x}}_{k+1}$, as well as the state covariance estimate $\mathbf{\hat{P}}_{k+1}$ at time step $k+1$, can be predicted by summing up all weighted sampling points. Applying the observation equation $h(\cdot)$ at first, the sampling points of the output vector $\mathbf{\hat{Y}}_{k+1}^i$ at time $k+1$ can be found and finally weighted to the estimated output vector $\mathbf{\hat{y}}_{k+1}$ as well:\\
\begin{equation}
\mathbf{\hat{X}}_{k+1}^i=f\left(\mathbf{\tilde{X}}_k^i,\mathbf{u}_k\right)\label{eq:state}\\
\end{equation}
\begin{equation}
\mathbf{\hat{x}}_{k+1}=\sum_{i=0}^{2n}W_m^i\mathbf{\hat{X}}_{k+1}^i\\
\end{equation}
\begin{equation}
\mathbf{\hat{P}}_{k+1}=\sum_{i=0}^{2n}W_c^i(\mathbf{\hat{X}}_{k+1}^i-\mathbf{\hat{x}}_{k+1})(\mathbf{\hat{X}}_{k+1}^i-\mathbf{\hat{x}}_{k+1})^T+\mathbf{Q}\\
\end{equation}
\begin{equation}
\mathbf{\hat{Y}}_{k+1}^i=h\left(\mathbf{\hat{X}}_{k+1}^i,\mathbf{u}_{k+1}\right)\\\label{eq:Measurement}
\end{equation}
\begin{equation}
\mathbf{\hat{y}}_{k+1}=\sum_{i=0}^{2n}W_m^i\mathbf{\hat{Y}}_{k+1}^i\\
\end{equation}

On basis of the innovation error $\mathbf{e}_{k+1}=\mathbf{y}_{k+1}-\mathbf{\hat{y}}_{k+1}$, the estimated state $\mathbf{\hat{x}}_{k+1}$ is corrected using the Kalman gain $\mathbf{K}_{k+1}$, which is calculated by the covariances $\mathbf{\hat{P}}_{yy,k+1}$ and $\mathbf{\hat{P}}_{xy,k+1}$. Finally it yields the predicted and corrected state $\mathbf{\tilde{x}}_{k+1}$ and covariance $\mathbf{\tilde{P}}_{k+1}$ respectively:
\begin{equation}
\mathbf{\hat{P}}_{yy,k+1}=\sum_{i=0}^{2n}W_c^i(\mathbf{\hat{Y}}_{k+1}^i-\mathbf{\hat{y}}_{k+1})(\mathbf{\hat{Y}}_{k+1}^i-\mathbf{\hat{y}}_{k+1})^T+\mathbf{R}\label{P_yy}\\
\end{equation}
\begin{equation}
\mathbf{\hat{P}}_{xy,k+1}=\sum_{i=0}^{2n}W_c^i(\mathbf{\hat{X}}_{k+1}^i-\mathbf{\hat{x}}_{k+1})(\mathbf{\hat{Y}}_{k+1}^i-\mathbf{\hat{y}}_{k+1})^T\\
\end{equation}
\begin{equation}
\mathbf{K}_{k+1}=\mathbf{P}_{xy,k+1}\mathbf{P}_{yy,k+1}^{-1}\\
\end{equation}
\begin{equation}
\mathbf{\tilde{x}}_{k+1}=\mathbf{\hat{x}}_{k+1}+\mathbf{K}_{k+1}\mathbf{e}_{k+1}=\mathbf{\hat{x}}_{k+1}+\mathbf{K}_{k+1}(\mathbf{y}_{k+1}-\mathbf{\hat{y}}_{k+1})\\
\end{equation}
\begin{equation}
\mathbf{\tilde{P}}_{k+1}=\mathbf{\hat{P}}_{k+1}-\mathbf{K}_{k+1}\mathbf{\hat{P}}_{yy,k+1}\mathbf{K}_{k+1}^T\\
\end{equation}

For the parameter estimation, the previous state vector $\mathbf{x}_k$, including displacement, velocity or acceleration information, has to be augmented by a parameter vector $\boldsymbol{\theta}_k$, containing all to be identified parameters. It yields the augmented state vector $\mathbf{x}_k^a$. Afterwards the state equation has to be changed and the UKF can be applied using $\mathbf{x}_k^a$.
\begin{equation}
\mathbf{x}_k^a=\begin{bmatrix}
\mathbf{x}_k\\ \boldsymbol{\theta}_k
\end{bmatrix}
\end{equation}

\subsection{Adaptation scheme for the system identification method}\label{sec2.2}

For systems with both initial nonlinearities and abrupt changes, an adaptation procedure is presented as follows. As the corrected state covariance $\mathbf{\tilde{P}}_k$ describes the confidence of the estimated and corrected state $\mathbf{\tilde{x}}_{k}$, it can be used to influence the upcoming parameter estimation step, i.e. a high state covariance yield more sensitive system identification and nonlinearities can be identified better.

For this purpose, firstly, similar to Bisht and Singh\cite{Bisht.2014}, the trigger parameter $\gamma$ based on the innovation error $\mathbf{e}_{k+1}$ is introduced. In contrast to Bisht and Singh, the innovation error is normalized by the measurement noise covariance $\mathbf{R}$ instead of the measurement covariance $\mathbf{P}_{yy}$ allowing the trigger parameter $\gamma$ to be independent from the measurement noise level:
\begin{equation}
\gamma=\mathbf{e}_{k+1}^T\mathbf{R}^{-1}\mathbf{e}_{k+1}\label{eq:adaption}
\end{equation}

For $m$ sensors with the identical constant measurement noise covariance $R_i=R$, $\gamma$ reads:
\begin{equation}
\gamma=\begin{bmatrix}e_1\\\vdots\\e_m\end{bmatrix}^T\begin{bmatrix}R_1&\cdots&0\\\vdots&\ddots&\vdots\\0&\cdots&R_m\end{bmatrix}^{-1}\begin{bmatrix}e_1\\\vdots \\e_m\end{bmatrix}=\frac{e_1^2}{R_1}+\cdots+\frac{e_m^2}{R_m}=\frac{e_1^2+\cdots+e_m^2}{R}\label{eq:adaption2}.
\end{equation}

To detect system changes, a threshold $\gamma_0$ is defined as an adaptation criterion, which has to be exceeded by $\gamma$ in adaptation cases:
\begin{equation}
\gamma\geqslant\gamma_0 \Rightarrow \text{adaptation}
\end{equation}

Bisht and Singh\cite{Bisht.2014} proposed to choose a constant threshold based on the \textit{a posteriori} known covariance of the measurement signal $\mathbf{y}_k$, whereas Rahimi \textit{et al.}\cite{Rahimi.2017} compensated the unknown \textit{a posteriori} information by introducing variable thresholds over time. However, different than the previous approaches, the in this paper proposed threshold avoids both the necessity of \textit{a posteriori} knowledge as well as the high sensitivity resulting from variable thresholds. On basis of known sensor numbers $m$ and measurement noise covariance $R$, the threshold $\gamma_0$ is calculated.

\begin{figure}[!]
	\begin{center}
		\includegraphics[width=\textwidth]{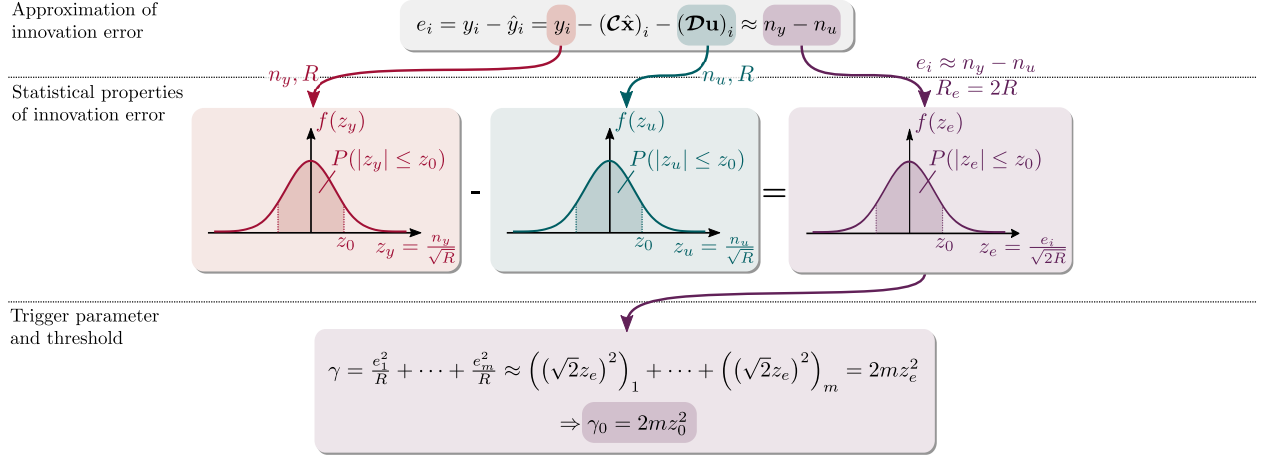}
	\end{center}
	\caption{Proposed scheme for the determination of the adaptation threshold $\gamma_0$ applied for accelerometers.}
	\label{A_UKF_Probability}
\end{figure}

Figure \ref{A_UKF_Probability} shows the individual steps of the derivation of the threshold, including the approximation of the innovation error with noise terms, the corresponding statistics, the realization of the trigger parameter $\gamma$ and the threshold $\gamma_0$.

The innovation error $\mathbf{e}_{k+1}$ contains both the system and measurement errors. System changes correspond to system errors. For the identification of system changes from the innovation error, the threshold $\gamma_0$ must cover with a certain probability the measurement error portion of the innovation error. If the trigger parameter $\gamma$ exceeds this threshold $\gamma_0$, the existence of a system error, i.e. system change, is ensured.

For the definition of such a threshold, we consider firstly a constant system behavior without any changes, i.e. no system errors. In this case, the innovation error can be solely approximated by the measurement error, Figure \ref{A_UKF_Probability} (Step 1: Approximation of innovation error). For the sake of simplicity, the time steps are not explicitly given for each parameter, since each parameter corresponds to the same time step. The measurement error is computed from the difference between the true measurement signal $y_i$ and the predicted measurement signal $\hat{y}_i$. The predicted measurement signal is calculated by the observation equation by the output matrix $\boldsymbol{\mathcal{C}}$ with the predicted state $\hat{\mathbf{x}}$ and the transition matrix $\boldsymbol{\mathcal{D}}$ with the input $\mathbf{u}$ (Section \ref{sec2.4}). The predicted state is independent from the measurement error. Accordingly, output $n_y$ and the input $n_u$ govern the measurement error. If the system motion is observed by displacement and velocity sensors, the transition matrix $\boldsymbol{\mathcal{D}}$ becomes zero, so that the innovation error $e_i$ solely depends on the output measurement noise $n_y$. If acceleration sensors are used, $\boldsymbol{\mathcal{D}}$ is an identity matrix and, consequently, $e_i$ is approximated by the difference of both the output $n_y$ and the input $n_u$ measurement noises.

Both measurement noises are assumed to be Gaussian and each has a covariance of R. Accordingly, their superposition can be treated as Gaussian as well \cite{BarShalom.2001}, Figure \ref{A_UKF_Probability} (Step 2: Statistical properties of innovation error). Consequently, $e_i$ has a mean of zero and its variance $R_e$ can be written as the sum of both variances as $2R$. The innovation error $e_i$ can now be expressed for each sensor by a standard normally distributed variable $z_e$ instead of $e_i$ and $R$.

As shown in Figure \ref{A_UKF_Probability} (Step 3: Trigger parameter and threshold), substituting $z_e$ instead of $e_i$ and $R_i$ in Equation 19 yields for the case of accelerometers $\gamma\approx2mz_e^2$, which solely depends on the number of sensors $m$, the variable $z_e$ and the scalar $2$, which results from the choice of accelerometers. The scalar changes to $\gamma\approx mz_e^2$ for displacement and velocity sensors. Consequently, a parameter $\delta$ is introduced in the calculation of the trigger parameter as

\begin{equation}
\gamma\approx\delta m z_e^2
\end{equation}

\noindent with $\delta=1$ for displacement and velocity sensors and $\delta=2$ for accelerometers respectively. Accordingly, the corresponding threshold is given by

\begin{equation}
\label{eq:threshold}
\gamma_0=\delta mz_0^2.
\end{equation}

Now, since $\delta$ and $m$ are system dependent preset parameters, $z_0$ governs the threshold based on the exceeding probability of the Gaussian distribution. For the variable $z_0=3\sqrt{2}$, which corresponds to an exceeding probability of $\SI{99.998}{\%}$\cite{Bendat.2010}, the threshold yields $\gamma_0=72$ for two accelerometers. This threshold value will be used in the performance studies in Section \ref{sec3.1}. The presented threshold, Equation \ref{eq:threshold}, is valid for monitoring systems consisting of either only displacement and velocity sensors or only accelerometers. Considering mixed sensor types in the monitoring system, instead, the threshold has to be derived individually as shown above.

\begin{figure}[!]
	\begin{center}
		\includegraphics[width=1\textwidth]{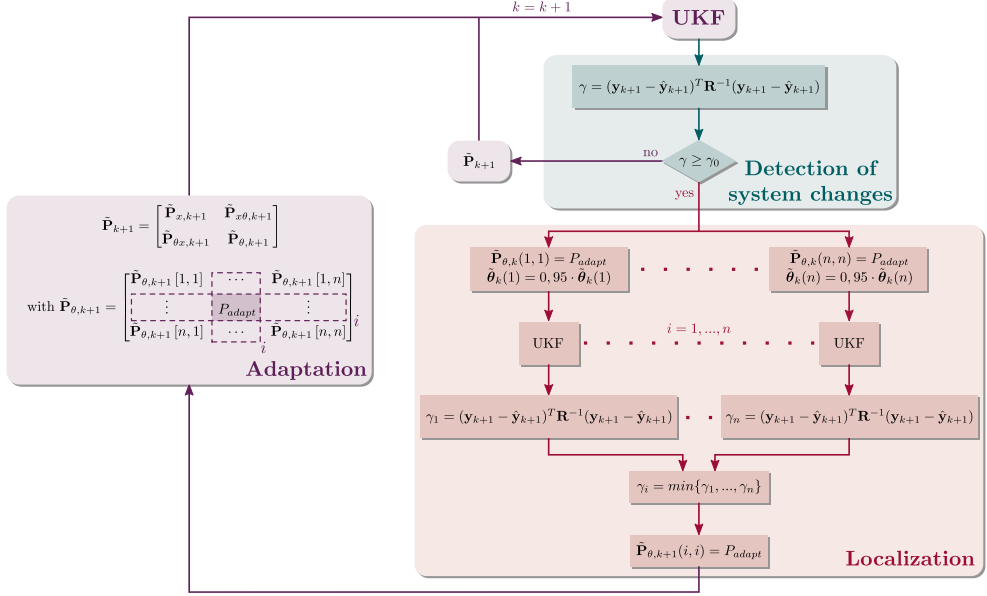}
	\end{center}
	\caption{Detection and localization of system changes.}
	\label{A_UKF}
\end{figure}
After detecting abrupt system changes, a localization algorithm has to follow. For this purpose, the localization scheme of Bisht and Singh\cite{Bisht.2014} is extended as shown in Figure \ref{A_UKF}. The flowchart of the proposed A-UKF presents besides the detection of system changes the adaptation step, in particular, consisting of localization and covariance adaptation. 
In the following paragraph the subscript $\theta$ denotes covariances $P$, which are dependent on the system parameters $\theta$ only, and subscript $x$ analogously denotes the state dependent covariances only. To localize system changes, an additional UKF estimation step is shown in Figure \ref{A_UKF} for the next time step $k+1$. The state covariance component  $\tilde{\mathbf{P}}_{\theta,k+1}\left\lbrack i,i\right\rbrack$ is set to $P_{adapt}$ for each $i=1,...,n$ individually, where $P_{adapt}$ is a high constant covariance value, which is introduced to increase the sensitivity of the parameter identification. Since only stiffness degradations are expected, each parameter $\tilde{\theta}_i$ with corresponding index $i$ is additionally decreased by \SI{5}{\%}, different than previous studies, in order to facilitate the localization. Accordingly, for each index $i$ now a different set of $\tilde{\mathbf{P}}_{k+1}$ and $\boldsymbol{\tilde{\theta}}_{k+1}$ exists.
For each of these sets and otherwise unchanged conditions a single calculation step of the UKF is executed and finally the trigger parameter $\gamma_i$ of Equation \ref{eq:adaption} is recalculated. Now assuming, that the lowest value of $\gamma_i$ describes the lowest system error and, thus, yields the best estimate for the system properties, the related index $i$ belongs to the degrading parameter $\theta_i$. For the next simulation step $k+1$ solely the state covariance component $\tilde{\mathbf{P}}_{\theta,k+1}\left\lbrack i,i\right\rbrack$ of the localized index $i$ is substituted by the new state covariance value $P_{adapt}$, which has to be chosen in advance and is highly dependent on the chosen system noise covariance $\mathbf{Q}$ and the present measurement noise covariance $\mathbf{R}$. The parameter has to be chosen as high as possible to enable a system identification of abrupt changes. Section \ref{sec3.3} will give a detailed simulation example of how to choose $P_{adapt}$.

\subsection{Application of the system identification method on MDoF+TMD systems}\label{sec2.3}

\begin{figure}[!]
	\begin{center}
		\includegraphics[width=1\textwidth]{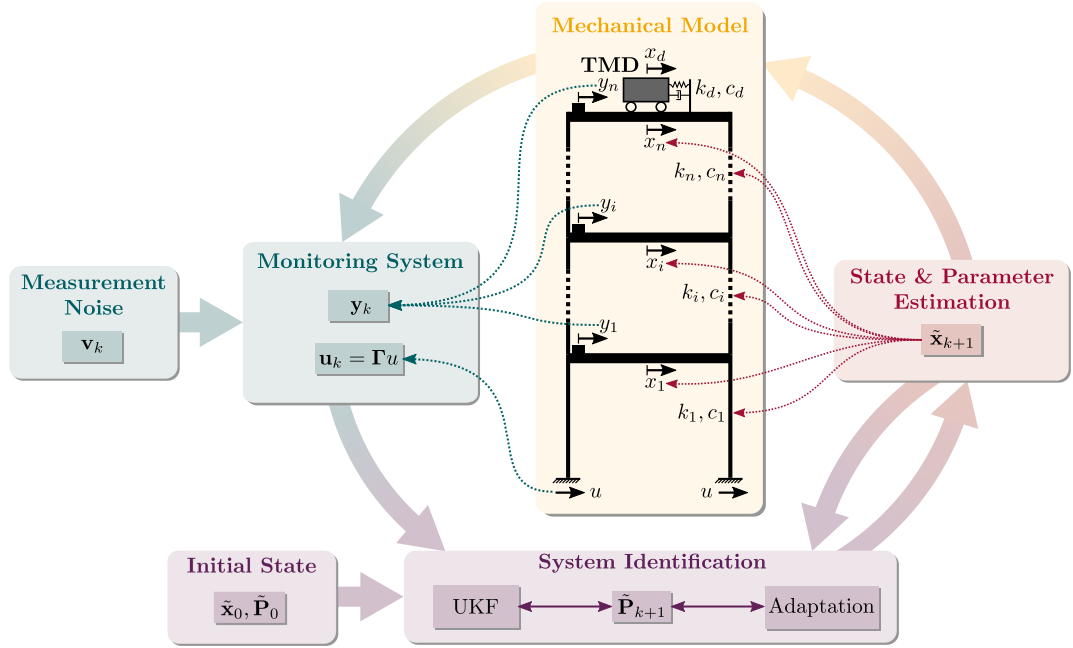}
		\caption{System identification scheme for MDoF+TMD systems.}
		\label{Flowchart}
	\end{center}
\end{figure}
The theory is introduced using the example case, at which a TMD is attached at the top DoF of a MDoF frame structure, Figure \ref{Flowchart}. The structure is instrumented with a monitoring system and the proposed system identification method will be implemented on this system to obtain its abrupt stiffness changes.
Thus the equation of motion with stiffness, damping and mass matrices $(\mathbf{K},\mathbf{C},\mathbf{M})$ can be set up for a seismic ground excitation distributed equally over the height of the system as
\begin{equation}
\mathbf{M}\mathbf{\ddot{x}}(t)+\mathbf{C}\mathbf{\dot{x}}(t)+\mathbf{K}(t)\mathbf{x}(t)=\mathbf{M}\mathbf{\Gamma}u(t)\hspace{10mm}\text{with}\ \mathbf{\Gamma}=\begin{bmatrix}1 & \cdots & 1\end{bmatrix}.
\end{equation}

The stiffness matrix $\mathbf{K}$ is assumed to be time variant with abrupt changes assembling a nonlinear structural behavior. Whereas the to be identified stiffness parameters $k_1(t),...,k_n(t)$ are time variant, the damping constants $c_1,...,c_n$ and masses $m_1,...,m_n$ remain constant during the simulation as well as the initially adjusted TMD parameters $k_d,c_d,m_d$.

A monitoring system is set up to observe the actual system responses. As it is shown in Figure \ref{Flowchart}, sensors are assumed to be placed on each DoF $i=1,...,n$. However, for other systems, e.g. high-rise structures with a large number of DoFs, a different sensor layout with a reduced amount of sensors is possible with a reduced accuracy. For the monitoring system all motion sensors (e.g. displacement, velocity, acceleration, forces) are possible. In the scope of this paper, accelerometers are used only, since they are the most commonly used sensor types for vibration measurements. Sensor properties, such as offset and RMS-value of the measurement noise are required for later system identification steps and the adaptation step, in particular. Using the response $\mathbf{y}$ and input signals $\mathbf{u}$, the in Section \ref{sec2.1} and \ref{sec2.2} introduced adaptive system identification method is applied for the joint state and parameter estimation computing the corrected and estimated state vectors, consisting of displacements, velocities and system stiffnesses.

The challenge for identification of highly damped systems (e.g. TMD) is to deal with rapidly decreasing vibration amplitudes compared to lightly damped systems. For such a system, a system identification is, therefore, only possible during a significantly smaller time period. In particular, for strongly (non-classically) damped systems, special attention has to be paid on the sensitivity or filter settings, respectively, of the system identification as well as the used mathematical models. This aspect will be elaborated in Section \ref{sec3} by three parametric studies on a MDoF structure with and without TMD.

\subsection{Modelling of the MDoF+TMD systems}\label{sec2.4}

As described in Section \ref{sec2.1} the UKF requires a state equation $f(\cdot)$ and an observation equation $h(\cdot)$, which are herein assumed as stepwise linear state-space representations. Starting with the state equation, the equation of motion of the previously in Section \ref{sec2.3} described system can be rewritten to a differential equation of $1^{st}$ order as follows:
\begin{equation}
\mathbf{\dot{x}}(t)=\boldsymbol{\mathcal{A}}(t)\mathbf{x}(t)+\boldsymbol{\mathcal{B}}\mathbf{u}(t)+\mathbf{w}(t)\label{state eq.}
\end{equation}

The system matrix $\boldsymbol{\mathcal{A}}(t)$ and input matrix $\boldsymbol{\mathcal{B}}$ contain the nonlinear system properties and information of input signals, respectively. An additive noise $\mathbf{w}(t)$ is added to the state equation describing the system noise, including errors of the mathematical model.

The monitoring system is transferred to the observation equation, where the output matrix $\boldsymbol{\mathcal{C}}$ and transition matrix $\boldsymbol{\mathcal{D}}$ describe the sensor layout of number, type and position. The result is finally enhanced by the noise component $\mathbf{v}(t)$, representing measurement noise:
\begin{equation}
\mathbf{y}(t)=\boldsymbol{\mathcal{C}}\mathbf{x}(t)+\boldsymbol{\mathcal{D}}\mathbf{u}(t)+\mathbf{v}(t)\label{observation eq.}\\
\end{equation}

Both state and measurement equations are given in continuous time so far. However, the system identification method requires a discrete time formulation, since the measurement data has a discrete form. A discretization can be realized by many methods, e.g. Euler or $4^{th}$ order Runge Kutta method. Although every method has different characteristics and calculation rules, all of them can be compared by the order of convergence $p$, defined by discretization errors. Higher orders of convergences generally yield more accurate results, but also have higher computational costs. In case of the explicit Euler method the order of convergence is $p=1$ and for $4^{th}$ order Runge Kutta $p=4$, respectively. In this paper, however, the discretization is done by a Taylor expansion developed from the analytical solution with orders of convergence $p=1-4$. This approach is preferred here, since all $p=1-4$ easily can be implemented based on one model only allowing a parametric study of the influence of model accuracy, Section \ref{sec3.4}. The discretization yields the matrices $\boldsymbol{\mathcal{A}}_d$ and $\boldsymbol{\mathcal{B}}_d$:
\begin{equation}
\boldsymbol{\mathcal{A}}_d=e^{\boldsymbol{\mathcal{A}}T_s}=\sum_{i=0}^{\infty}\frac{1}{i!}\boldsymbol{\mathcal{A}}^iT_s^i\approx\mathbf{I}+\boldsymbol{\mathcal{A}}T_s+...+\frac{1}{p!}\boldsymbol{\mathcal{A}}^pT_s^p
\end{equation}
\begin{align}
\boldsymbol{\mathcal{B}}_d=\int_{0}^{T_s}e^{\boldsymbol{\mathcal{A}}\tau}\boldsymbol{\mathcal{B}}d\tau&=\sum_{i=0}^{\infty}\frac{1}{(i+1)!}\boldsymbol{\mathcal{A}}^i\boldsymbol{\mathcal{B}}T_s^{i+1}\notag\\&\approx\mathbf{0}+\boldsymbol{\mathcal{B}}T_s+...+\frac{1}{(p+1)!}\boldsymbol{\mathcal{A}}^p\boldsymbol{\mathcal{B}}T_s^{p+1}
\end{align}

Using the above discretization, the state and observation equation can be easily transformed into the discrete domain assuming real sampling, with the sampling time $T_s$:
\begin{equation}
\mathbf{x}_{k+1}=\boldsymbol{\mathcal{A}}_{d,k}\mathbf{x}_k+\boldsymbol{\mathcal{B}}_d\mathbf{u}_k+\mathbf{w}_k\label{state eq.}\\
\end{equation}
\begin{equation}
\mathbf{y}_k=\boldsymbol{\mathcal{C}}\mathbf{x}_k+\boldsymbol{\mathcal{D}}\mathbf{u}_k+\mathbf{v}_k\label{observation eq.}\\
\end{equation}

%

\section{Performance Studies}\label{sec3}

In this section, investigations on a two-degree-of-freedom (2-DoF) structure with and without TMD will be presented under seismic, white noise and impulse excitations. Detailed parameter studies are done regarding the system accuracy and convergence behavior of the system identification method considering TMD influence and abrupt stiffness changes of the structure. Recommendations to the filter and model setup are given for the investigated systems.

\subsection{Description of the investigated MDoF+TMD systems}\label{sec3.1}
\begin{figure}[!]
	\begin{center}
		\includegraphics[width=\textwidth]{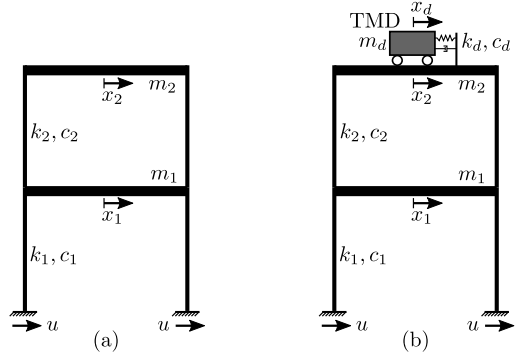}\\
	\end{center} 	
	\caption{Investigated systems: (a) 2-DoF and (b) 2-DoF+TMD.}
	\label{Systems}
\end{figure}

Two different systems are investigated: solely a 2-DoF structure as well as the same 2-DoF structure with a TMD attached at the top DoF, Figure \ref{Systems}. Stiffness, damping and mass matrices $(\mathbf{K}, \mathbf{C}, \mathbf{M})$ can be set up corresponding to the system parameters, listed in Table \ref{system properties}, as
\begin{gather}
\mathbf{K}=\begin{bmatrix} k_1+k_2 & -k_2 & 0 \\ -k_2 & k_2+k_d & -k_d \\ 0 & -k_d & k_d\end{bmatrix};\hspace{10mm}
\mathbf{C}=\begin{bmatrix} c_1+c_2 & -c_2 & 0 \\ -c_2 & c_2+c_d & -c_d \\ 0 & -c_d & c_d\end{bmatrix};\hspace{10mm}
\mathbf{M}=\begin{bmatrix} m_1 & 0 & 0 \\ 0 & m_2 & 0 \\ 0 & 0 & m_d\end{bmatrix}.
\end{gather}

The masses $(m_1, m_2)$ and the damping constants $(c_1, c_2)$ are time invariant values. The stiffness values $(k_1, k_2)$ consider abrupt system changes of $\SI{10}{\%}$ at each DoF corresponding to a $\SI{5}{\%}$ decrease of the $1^{st}$ natural frequency, which are realistic values to observe during load cases, such as earthquakes. The initial stiffness values are chosen as $k_1=\SI{12}{kN/m}$ and $k_2=\SI{10}{kN/m}$. For each load case, abrupt stiffness changes appear at a defined time step $t_f$ according to high interstory drifts between the individual DoFs.

\begin{table}
	\caption{System parameters of (a) 2-DoF and (b) 2-DoF+TMD.}
	\label{system properties}
	\begin{center}
		\begin{tabular}{ccccc}
			\toprule
			\textbf{System parameter} & \multicolumn{2}{c}{\textbf{(a) 2-DoF}} & \multicolumn{2}{c}{\textbf{(b) 2-DoF+TMD}}\\
			\cmidrule{2-3}\cmidrule{4-5}
			&$t\le t_f$&$t>t_f$&$t\le t_f$&$t>t_f$\\
			\midrule
			$m_1$ &\multicolumn{2}{c}{\SI{1}{t}}&\multicolumn{2}{c}{\SI{1}{t}}\\
			$m_2$ &\multicolumn{2}{c}{\SI{1}{t}}&\multicolumn{2}{c}{\SI{1}{t}}\\
			$m_d$ &\multicolumn{2}{c}{-}&\multicolumn{2}{c}{\SI{0.1}{t}}\\
			$k_1(t)$ &\SI{12}{kN/m}&\SI{10.8}{kN/m}& \SI{12}{kN/m}&\SI{10.8}{kN/m}\\
			$k_2(t)$ &\SI{10}{kN/m}&\SI{9}{kN/m}& \SI{10}{kN/m}&\SI{9}{kN/m}\\
			$k_d$ &\multicolumn{2}{c}{-}&\multicolumn{2}{c}{\SI{0.36}{kN/m}}\\
			$c_1$ &\multicolumn{2}{c}{\SI{0.1}{kNs/m}}&\multicolumn{2}{c}{\SI{0.1}{kNs/m}}\\
			$c_2$ &\multicolumn{2}{c}{\SI{0.1}{kNs/m}}&\multicolumn{2}{c}{\SI{0.1}{kNs/m}}\\
			$c_d$ &\multicolumn{2}{c}{-}&\multicolumn{2}{c}{\SI{0.051}{kNs/m}}\\
			$f_1$ &\multicolumn{2}{c}{\SI{0.33}{Hz}}&\multicolumn{2}{c}{\SI{0.27}{Hz}}\\
			$f_2$ &\multicolumn{2}{c}{\SI{0.84}{Hz}}&\multicolumn{2}{c}{\SI{0.36}{Hz}}\\
			$f_3$ &\multicolumn{2}{c}{-}&\multicolumn{2}{c}{\SI{0.84}{Hz}}\\
			\bottomrule
		\end{tabular}
	\end{center}
\end{table}
The attached TMD is defined by the time invariant parameters $k_d, c_d, m_d$, which are chosen in the initial time step. The undamaged 2-DoF structure has a natural frequency of $f_1=\SI{0.33}{Hz}$ and a damping ratio of $D_1=\SI{0.92}{\%}$ for the $1^{st}$ eigenmode, and analogously for the $2^{nd}$ eigenmode $f_2=\SI{0.84}{Hz}$ and $D_2=\SI{2.49}{\%}$. Table \ref{system properties} provides the remaining natural frequencies of the 2-DoF+TMD system. To adjust the damper parameters, several possible approaches are proposed in the literature. In this paper, we focus on the system identification and use the classical approach of Warburton \cite{Warburton.1982}. Assuming the structure to be lightly damped $(D_1=\SI{0.92}{\%})$ an application of Warburton is reasonable. The TMD is tuned to the $1^{st}$ natural frequency of the 2-DoF structure. Therefore, the mass ratio $\mu$, describing the relation of damper mass $m_d$ and generalized mass of the $1^{st}$ mode $\hat{m}_1$, the optimal damper frequency $f_{opt}$, dependent on $f_1$ and $\mu$, and finally the optimal damping ratio $D_{opt}$, dependent on $\mu$, are calculated:
\begin{gather}
\mu=\frac{m_d}{\hat{m}_1}=0.076\hspace{10mm}
f_{opt}=f_1\frac{\sqrt{1-\frac{\mu}{2}}}{1+\mu}=\SI{0.30}{Hz}\hspace{10mm}
D_{opt}=\sqrt{\frac{\mu(1-\frac{\mu}{4})}{4(1+\mu)(1-\frac{\mu}{2})}}=\SI{13.42}{\%}
\end{gather}
Subsequently, all damper parameters can be calculated using fundamental SDoF relations, Table \ref{system properties}.

In a final step, $\mathbf{K}$, $\mathbf{C}$ and $\mathbf{M}$ are transformed into the state-space representation. The time variant system matrix $\boldsymbol{\mathcal{A}}(t)$ and the input matrix $\boldsymbol{\mathcal{B}}$ can be set up according to the nonlinear system properties. We formulate the representation for a ground acceleration $\ddot{x}_g$ as input. Furthermore, the output matrix $\boldsymbol{\mathcal{C}}$ and transition matrix $\boldsymbol{\mathcal{D}}$ can be calculated as follows, describing a monitoring system of two accelerometers on both DoFs $x_1$ and $x_2$ and one accelerometer for the ground motion $\ddot{x}_g$:
\begin{gather}
\boldsymbol{\mathcal{A}}(t)=\begin{bmatrix} \mathbf{0}_{3\times 3} & \mathbf{I}_{3\times 3} & \mathbf{0}_{3\times 2}\\ -\mathbf{M}^{-1}\mathbf{K}(t) & -\mathbf{M}^{-1}\mathbf{C} & \mathbf{0}_{3\times 2} \\ \mathbf{0}_{2\times 3} & \mathbf{0}_{2\times 3} & \mathbf{0}_{2\times 2} \end{bmatrix}; \hspace{5mm} \boldsymbol{\mathcal{B}}=\begin{bmatrix} \mathbf{0}_{3\times 3} \\ \mathbf{I}_{3\times 3} \\ \mathbf{0}_{2\times 3}\end{bmatrix};\notag\\
\boldsymbol{\mathcal{C}}=\begin{bmatrix}-\mathbf{M}^{-1}\mathbf{K} & \hspace{1mm}-\mathbf{M}^{-1}\mathbf{C} \end{bmatrix}; \hspace{5mm} \boldsymbol{\mathcal{D}}=\begin{bmatrix} \mathbf{I}_{2\times 2} \end{bmatrix}
\end{gather}
with the input vector $\mathbf{u}$ and the output vector $\mathbf{y}$:
\begin{equation}
\mathbf{u}=\begin{bmatrix}	\ddot{x}_g\\ \ddot{x}_g\\ \ddot{x}_g\\	\end{bmatrix};\hspace{5mm}
\mathbf{y}=\begin{bmatrix}	\ddot{x}_1\\ \ddot{x}_2\\ \end{bmatrix}\\
\end{equation}

Both state and observation equations are calculated for the joint state and parameter estimation, i.e. the state vector $\mathbf{x}$ extends to an augmented state vector $\mathbf{x}^a$, including displacements, velocities and stiffnesses of the system:
\begin{equation}
\mathbf{x}^a=\begin{bmatrix}
x_1 & x_2 & x_d & \dot{x}_1 & \dot{x}_2 & \dot{x}_d & k_1 & k_2 \end{bmatrix}^T
\end{equation}

\begin{figure}
	\begin{center}
		\includegraphics[width=1\textwidth]{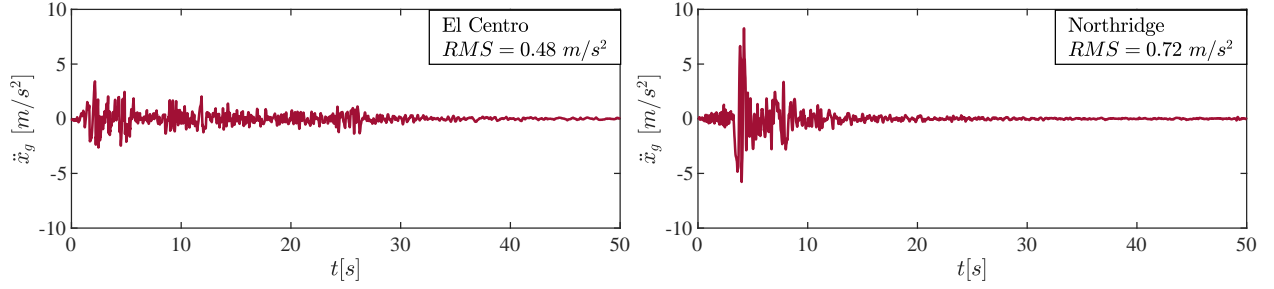}\\
	\end{center} 	
	\caption{Acceleration time histories of the El Centro (1940) (left) and the Northridge (1994) earthquakes (right).}
	\label{Fig:Earthquake_histories}
\end{figure}

For the investigations, the earthquake acceleration histories of the El Centro far field earthquake (1940) and the Northridge near field earthquake (1994) are considered, Figure \ref{Fig:Earthquake_histories}. Moreover, a white noise input with an RMS-value of $\SI{0.57}{m/s^2}$ and an impulse load of $\SI{80}{m/s^2}$ at $t=\SI{2}{s}$ are investigated. In all studies an additive white Gaussian noise of RMS$=\SI{0.01}{m/s^2}$, approximately $\SI{2}{\%}$ RMS-noise of the El Centro earthquake, is added to input as well as output measurement signals. The simulations are performed with a sampling time of $T_s=\SI{0.02}{s}$.

\subsection{Study 1: Identification of the augmented state including abrupt stiffness changes}\label{sec3.2}

The chosen filter setup of the identification algorithm applied herein is presented in Table \ref{filter_setup_1}. The state covariance $\mathbf{P}$ is chosen corresponding to the results of the parametric study 2, which will be presented in Section \ref{sec3.3}. According to the results of the parametric study, the new state covariance value after adaptation is chosen as high as possible as $P_{adapt}=10^0$. The initial covariance value, however, is chosen as low as possible $\mathbf{P}_0=10^{-6}\mathbf{I}_{8\times8}$, so that the identification algorithm does not show oversensitive reactions regarding the stiffness estimation before the abrupt stiffness change. The system noise covariance $\mathbf{Q}$ is chosen corresponding to the results of the parametric study 3, which will be presented in Section \ref{sec3.4}. Both 2-DoF and 2-DoF+TMD systems are modeled based on $3^{rd}$ order Taylor expansion. Effects of the used order for the Taylor expansion, in particular, will be also shown in Section \ref{sec3.4}. The measurement noise covariance $\mathbf{R}$ is calculated from the square of the RMS-value for the present noise, according to $\SI{2}{\%}$ RMS-noise of El Centro earthquake. The initial stiffness estimations $\hat{k}_1$ and $\hat{k}_2$ correspond to the real stiffness values $k_1$ and $k_2$, as shown in Table \ref{system properties}.

\begin{figure}
	\includegraphics[width=\textwidth]{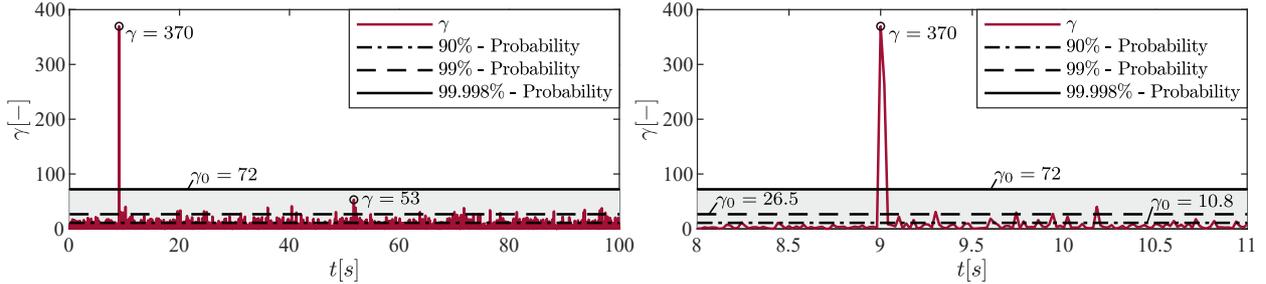}\\
	\caption{Study 1: Time history of the trigger parameter $\gamma$ (left). A time window from the time history around the abrupt stiffness change at $t=\SI{9}{s}$ (right).}
	\label{gamma}
\end{figure}

In Figure \ref{gamma} (left) the time history of the trigger parameter $\gamma$ is shown. The curve is calculated by the previously, in Section \ref{sec2.2}, introduced Equation \ref{eq:adaption}. In addition, the right diagram shows the time history for a time window around the abrupt stiffness change. The trigger parameter shows a peak value corresponding to the time step of the abrupt stiffness change $t=\SI{9}{s}$. After comparing the threshold $\gamma_0$, which is calculated from Equation \ref{eq:threshold}, the state covariance is adapted. In Figure \ref{gamma}, three different thresholds $\gamma_0=10.8$, $\gamma_0=26.5$ and $\gamma_0=72$ are shown according to the exceeding probability of $\SI{90}{\%}$, $\SI{99}{\%}$ and $\SI{99.998}{\%}$ respectively. For both probabilities $\SI{90}{\%}$ and $\SI{99}{\%}$ the threshold is exceeded several times with significantly high values (e.g. $\gamma=53$ at $t =\SI{52}{s}$). Best result is achieved with the threshold value of $\gamma_0=72$, which is exceeded only during the abrupt stiffness change.

\begin{figure}[!]
	\centering
	\includegraphics[width=0.98\textwidth]{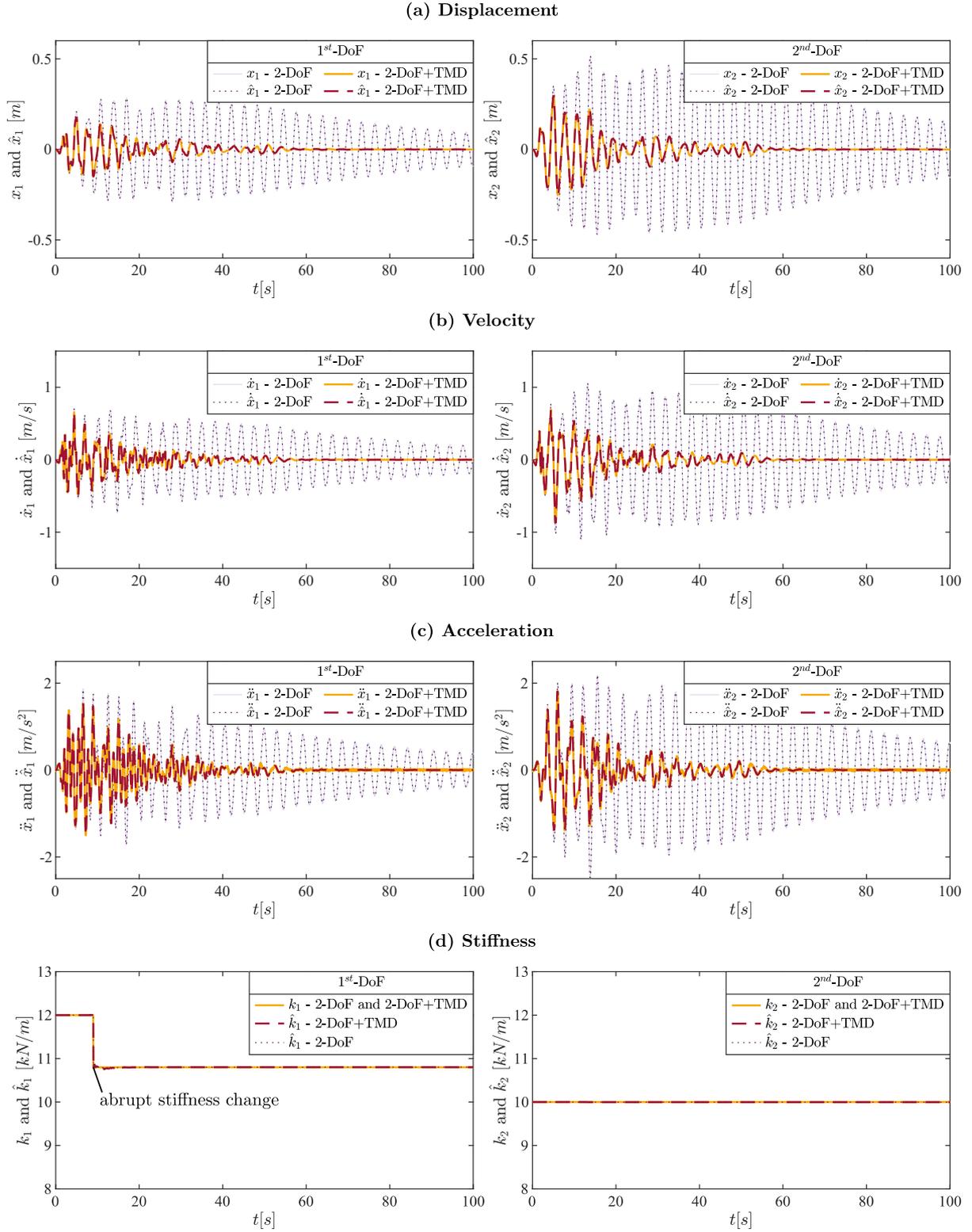}\\
	\caption{Study 1: Time histories of estimated $(\hat{\cdot})$ and true values of displacement (a), velocity (b), acceleration (c) and stiffness (d) of both DoFs during El Centro earthquake. Abrupt stiffness change at $1^{st}$ DoF.}
	\label{Fig:7_Study_1}
\end{figure}

In the first part of the study, we consider the El Centro earthquake excitation, including a stiffness degradation of $\SI{10}{\%}$ at the $1^{st}$ DoF after $t=\SI{9}{s}$ due to large story drift. Figure \ref{Fig:7_Study_1} compares the true values of the motion (displacement, velocity and acceleration) of both DoF of the structure and the stiffness time histories with those time histories, which are estimated by the proposed system identification method. Both cases with and without TMD are presented in the graphics. The abrupt stiffness change can be directly seen from the time histories of $k_1$ and $\hat{k}_1$. Both estimated and true values match with each other. The abrupt stiffness change is identified for both systems.

\begin{figure}[!]
	\centering
	\includegraphics[width=0.98\textwidth]{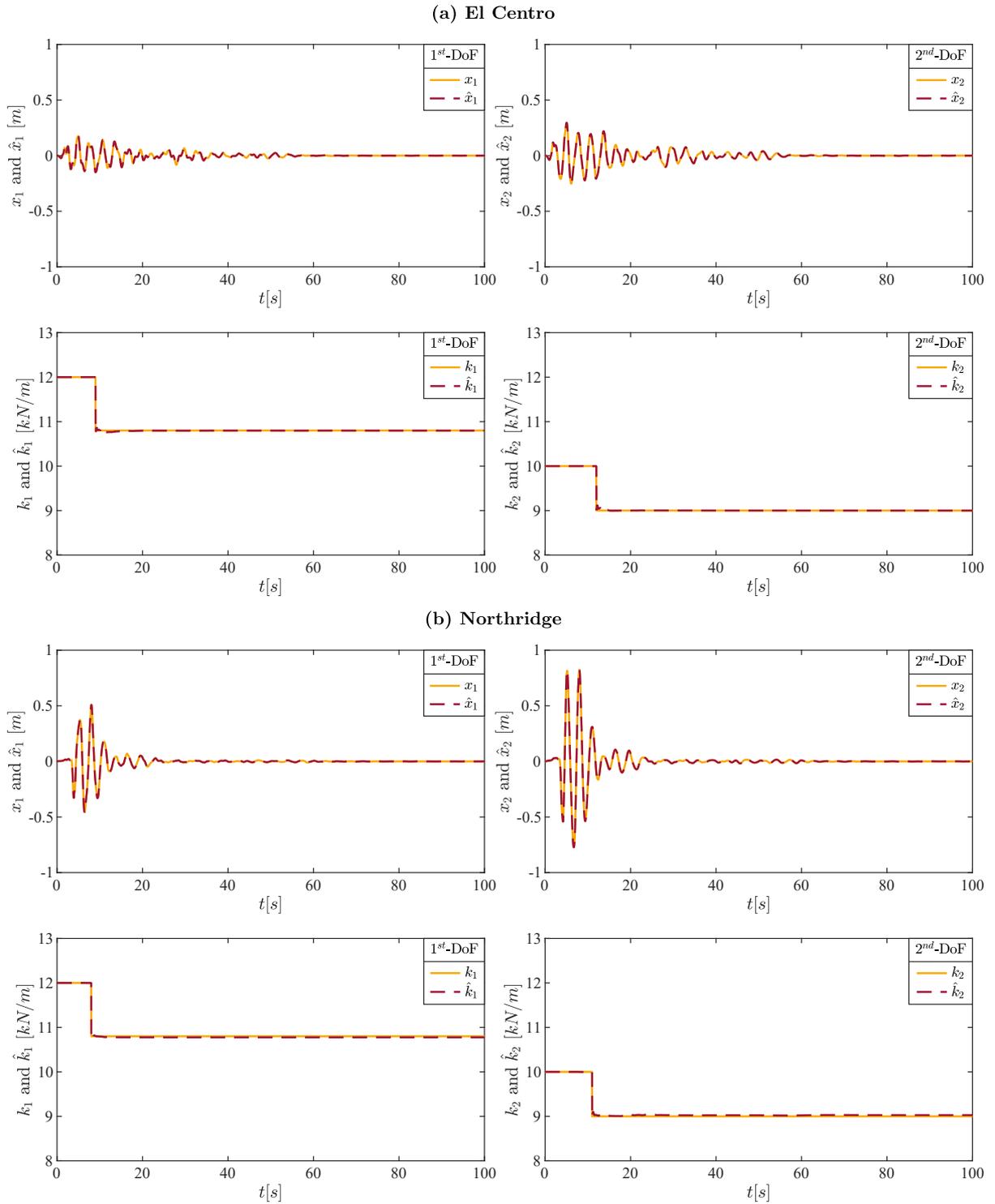}\\
	\caption{Study 1: Time histories of estimated $(\hat{\cdot})$ and true values of displacement and stiffness of both DoFs during El Centro (a) and Northridge (b) earthquake. Abrupt stiffness changes at both $1^{st}$ and $2^{nd}$ DoFs.}
	\label{Fig:8_Study_1}
\end{figure}

\begin{figure}[!]
	\centering
	\includegraphics[width=0.98\textwidth]{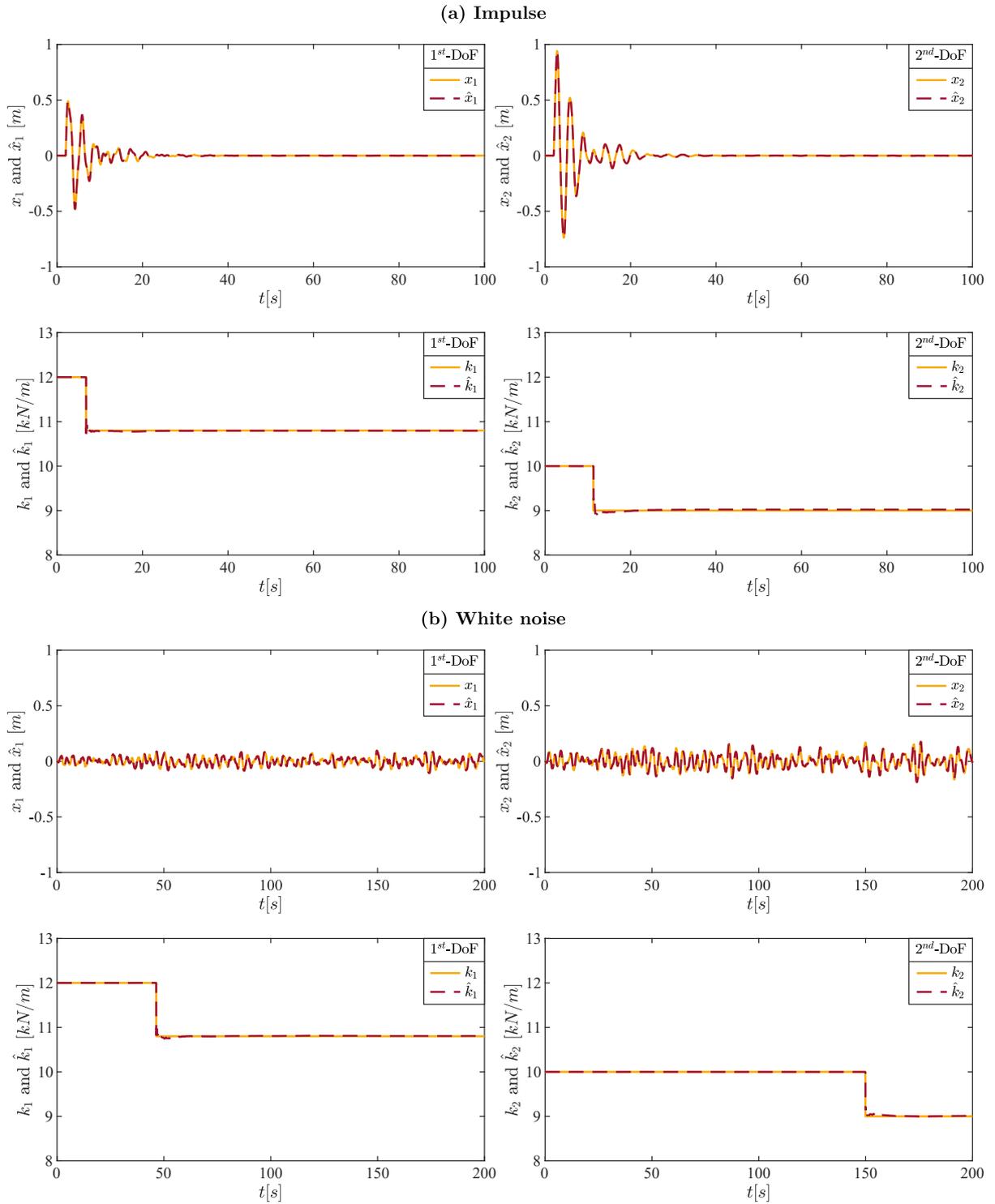}\\
	\caption{Study 1: Time histories of estimated $(\hat{\cdot})$ and true values of displacement and stiffness of both DoFs during impulse (a) and white noise (b) excitation. Abrupt stiffness changes at both $1^{st}$ and $2^{nd}$ DoFs.}
	\label{Fig:9_Study_1}
\end{figure}

In the second part of the study, we enhance our investigation by considering besides the both El Centro and Northridge earthquakes also impulse and white noise excitations. Furthermore, we allow an abrupt stiffness change on the $2^{nd}$ DoF as well. The occurrence times also in this second part of the study correspond to the interstory drift between $1^{st}$ DoF and $2^{nd}$ DoF. In Figures \ref{Fig:8_Study_1} and \ref{Fig:9_Study_1}, the corresponding time histories of the estimated and true values of the displacements and stiffness values are shown. A high accuracy of the estimated results is also observed here.

\begin{table}
	\caption{Study 1: Filter setup of the system identification method.}
	\label{filter_setup_1}
	\begin{center}
		\begin{tabular}{cccc}
			\toprule
			\textbf{Filter parameter}&\textbf{Value}&\textbf{Scaling factor}&\textbf{Value}\\
			\midrule
			$\mathbf{P}_0$&$10^{-6}\mathbf{I}_{8\times8}$&$\alpha$&$0.001$\\
			$P_{adapt}$&$10^{0}$&$\beta$&$2$\\
			$\mathbf{Q}$&$10^{-9}\mathbf{I}_{8\times8}$&$\kappa$&$0$\\
			$\mathbf{R}$&$10^{-4}\mathbf{I}_{3\times3}$&&\\
			\bottomrule
		\end{tabular}
	\end{center}
\end{table}

\subsection{Study 2: Effects of the state covariance}\label{sec3.3}

\begin{table}
	\caption{Study 2: Filter setup of the system identification method.}
	\label{filter_setup_2}
	\begin{center}
		\begin{tabular}{cccc}
			\toprule
			\textbf{Filter parameter}&\textbf{Value}&\textbf{Scaling factor}&\textbf{Value}\\
			\midrule
			$\mathbf{P}_0$&$10^{-8}\mathbf{I}_{8\times8}\cdots10^{0}\mathbf{I}_{8\times8}$&$\alpha$&$0.001$\\
			$\mathbf{Q}$&$10^{-9}\mathbf{I}_{8\times8}$&$\beta$&$2$\\
			$\mathbf{R}$&$10^{-4}\mathbf{I}_{3\times3}$&$\kappa$&$0$\\
			\bottomrule
		\end{tabular}
	\end{center}
\end{table}

The supplementary damping introduced by the TMD as well as abrupt stiffness changes of the structure shorten the time window, in which the proposed system identification method must complete its estimation. In this regard, the most powerful parameter is the state covariance $\mathbf{P}$. By increasing the state covariance, the reaction time of the identification method can be reduced. On the other hand, too high $P$ values can decrease the estimation accuracy. To clarify this effect, this study performs calculations with different constant $P$ values between $10^{-8}$ and $10^0$. Further filter parameters are shown in Table \ref{filter_setup_2}. Calculations are performed using $3^{rd}$ order Taylor expansion based models of 2-DoF and 2-DoF+TMD systems under the Northridge earthquake. The initial stiffness estimates of the structure are assigned as $\hat{k}_1=\SI{14.4}{kN/m}$ and $\hat{k}_2=\SI{12}{kN/m}$, which are $\SI{20}{\%}$ higher than the true stiffness values of $k_1=\SI{12}{kN/m}$ and $k_1=\SI{10}{kN/m}$.

\begin{figure}[!]
	\centering
	\includegraphics[width=\textwidth]{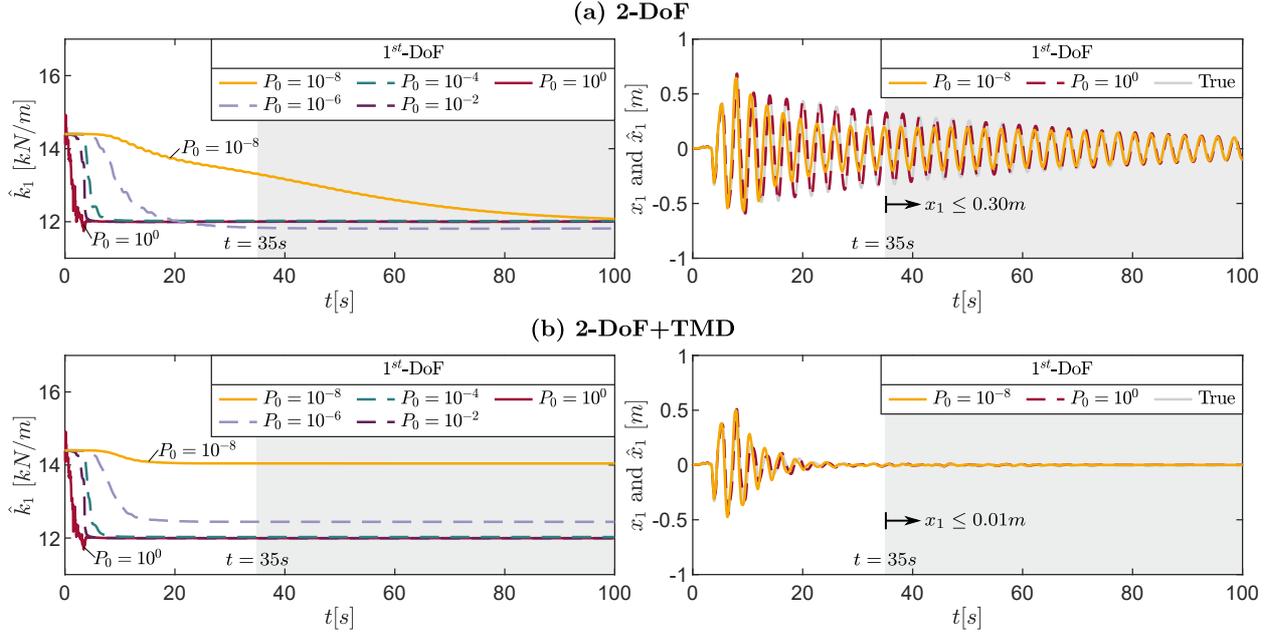}\\
	\caption{Study 2: Time histories of estimated stiffness $(\hat{k}_1)$ of $1^{st}$ DoF (left). Time histories of estimated $(\hat{x}_1)$ and true $(x_1)$ displacement of $1^{st}$ DoF (right). State covariance values $P_0$ are varying. Calculations are performed for (a) 2-DoF and (b) 2-DoF+TMD systems under Northridge earthquake.}
	\label{Fig:10_Study_2}
\end{figure}

%
Figure \ref{Fig:10_Study_2} (left) shows the true and estimated values of the $1^{st}$ DoF stiffness $k_1$ and $\hat{k}_1$. On the right side in Figure \ref{Fig:10_Study_2} we see the true and estimated values of the displacement of the $1^{st}$ DoF $x_1$ and $\hat{x}_1$. The displacement time histories are shown for the selected state covariance values of $10^{-8}$ and $10^0$. From the comparison of the displacement time histories the effect of the TMD can be clearly observed from the short vibration duration. Already after $\SI{35}{s}$ the vibration of the 2-DoF+TMD system is reduced below $\SI{0.01}{m}$. At the same time step, the vibration of the 2-DoF structure without TMD still continues with an amplitude of $\SI{0.30}{m}$. This difference governs the required accuracy level of the identification method.

In the time histories of the stiffness, we observe, in particular for lower $P$ values, that as soon as the vibrations vanish the estimated stiffness of the 2-DoF+TMD system converges to a constant value, which is far away from the real stiffness value. For instance, the estimated stiffness value of 2-DoF+TMD system is for $P=10^{-8}$ approximately $\SI{14}{kN/m}$, which does not match the true stiffness value of $\SI{12}{kN/m}$. For the same $P$ value of $10^{-8}$, the estimated stiffness of the 2-DoF structure without TMD converges slowly to the true stiffness value as the structure is still continuing to oscillate.

By increasing the $P$ value, we observe from the results that both systems can be identified with high accuracy. For the 2-DoF+TMD system, the correct stiffness value is estimated with $P=10^{-4}$. On the other hand, as stated before, the 2-DoF structure is estimated already with $P=10^{-8}$.
A further increase of the $P$ value causes the system identification method to behave oversensitive and the estimated stiffness course begins for both systems to fluctuate. With high $P$ values we observe at the beginning of the both time histories initially underestimated stiffness values.

Accordingly, the $P$ value must be chosen depending on the expected abrupt changes and the type of the system, which is a challenge for all UKF-based system identification methods. To overcome this effect, as introduced in Section \ref{sec2.2}, the proposed parameter identification algorithm tunes the state covariance in an adaptive manner.

\subsection{Study 3: Modeling effects}\label{sec3.4}

The accuracy of recursive system identification methods is directly related with the accuracy of the chosen mathematical model describing the system properties. The error inherent in the chosen mathematical model is considered in the proposed UKF-based identification method by the system noise covariance $\mathbf{Q}$. However, due to additional damping of TMDs, the accuracy sensitivity of the identification process increases. Therefore, $\mathbf{Q}$ struggles to realize the desired identification efficiency. Accordingly, the necessity of an accurate mathematical model increases for MDoF+TMD systems.

In this section, to show the modeling effect, four mathematical models are investigated using $\boldsymbol{\mathcal{A}}_d$ and $\boldsymbol{\mathcal{B}}_d$ discretization, introduced in Section \ref{sec2.4}, by Taylor expansions of $1^{st}$ to $4^{th}$ order of convergence, Table \ref{Table:discretized_matrices}. During the study, different $\mathbf{Q}$ matrices, which are constant over simulation time, are introduced varying from $10^{-8}\mathbf{I}_{8\times8}$ to $10^{-15}\mathbf{I}_{8\times8}$. Two load scenarios are investigated: The El Centro and the Northridge earthquakes. To determine the accuracy of the final stiffness estimation, the deviation parameter $\Delta k_i$ is introduced, which defines the percentage deviation of the final estimated stiffness $\hat{k}_i$ to the true value $k_i$:
\begin{equation}
\Delta k_i=\frac{\lvert k_i-\hat{k}_i\rvert}{k_{i}} \ \left[\%\right]
\end{equation}
\begin{table}
	\caption{Study 3: Discretizations of the system matrix $\boldsymbol{\mathcal{A}}_d$ and the output matrix $\boldsymbol{\mathcal{B}}_d$ with up to $4^{th}$ order Taylor expansion.}
	\label{Table:discretized_matrices}
	\begin{center}
		\begin{tabular}{ccccc}
			\toprule
			\textbf{Matrix}&\multicolumn{4}{c}{\textbf{Taylor expansions}}\\
			\cmidrule{2-5}
			&$p=1$&$p=2$&$p=3$&$p=4$\\
			\midrule
			$\boldsymbol{\mathcal{A}}_d$&$\mathbf{I}+\boldsymbol{\mathcal{A}}T_s$&$\frac{1}{2!}\boldsymbol{\mathcal{A}}^2T_s^2$&$\frac{1}{3!}\boldsymbol{\mathcal{A}}^3T_s^3$&$\frac{1}{4!}\boldsymbol{\mathcal{A}}^4T_s^4$\\
			$\boldsymbol{\mathcal{B}}_d$&$\boldsymbol{\mathcal{B}}T_s$&$\frac{1}{2!}\boldsymbol{\mathcal{A}}\boldsymbol{\mathcal{B}}T_s^2$&$\frac{1}{3!}\boldsymbol{\mathcal{A}}^2\boldsymbol{\mathcal{B}}T_s^3$&$\frac{1}{4!}\boldsymbol{\mathcal{A}}^3\boldsymbol{\mathcal{B}}T_s^4$\\
			\bottomrule
		\end{tabular}
	\end{center}
\end{table}
\begin{table}[bht]
	\caption{Study 3: Filter setup of the system identification method.}
	\label{filter_setup_3}
	\begin{center}
		\begin{tabular}{cccc}
			\toprule
			\textbf{Filter parameter}&\textbf{Value}&\textbf{Scaling factor}&\textbf{Value}\\
			\midrule
			$\mathbf{P}_0$&$\mathbf{I}_{8\times8}$&$\alpha$&$0.001$\\
			$\mathbf{Q}$&$10^{-8}\mathbf{I}_{8\times8}\dots10^{-15}\mathbf{I}_{8\times8}$&$\beta$&$2$\\
			$\mathbf{R}$&$10^{-4}\mathbf{I}_{3\times3}$&$\kappa$&$0$\\
			\bottomrule
		\end{tabular}
	\end{center}
\end{table}

In this study, the initial stiffness estimates are chosen to be $k_1=\SI{14.4}{kN/m}$ and $k_2=\SI{12}{kN/m}$, which are $\SI{20}{\%}$ higher than the true stiffness values of $k_1=\SI{12}{kN/m}$ and $k_1=\SI{10}{kN/m}$. Accordingly, a nonlinear parameter identification is required. Besides this fact, in this study, the structure is assumed to behave linearly during the earthquake excitation without any abrupt stiffness changes. All remaining filter setup parameters are shown in Table \ref{filter_setup_3}.

\begin{figure}[bht]
	\centering
	\includegraphics[width=\textwidth]{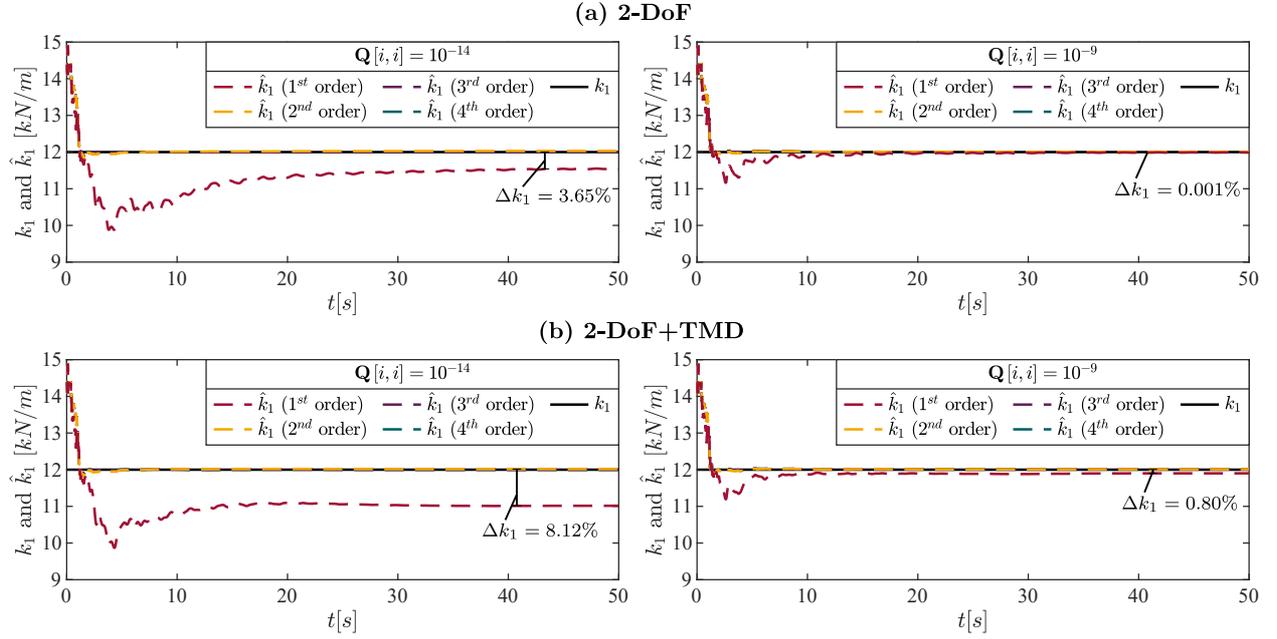}\\
	\caption{Study 3: Time histories of the estimated and true stiffness values $\hat{k}_1$ and $k_1$ for the selected system noise covariance levels $\mathbf{Q}\left\lbrack i,i\right\rbrack$ of $10^{-14}$ (left) and $10^{-9}$ (right). Estimations are calculated using $1^{st}$-$4^{th}$ order Taylor expansion discretizations for the (a) 2-DoF and (b) 2-DoF+TMD system under El Centro earthquake.}
	\label{Fig:11_Study_3}
\end{figure}
\begin{figure}[!htb]
	\centering
	\includegraphics[width=0.5\textwidth]{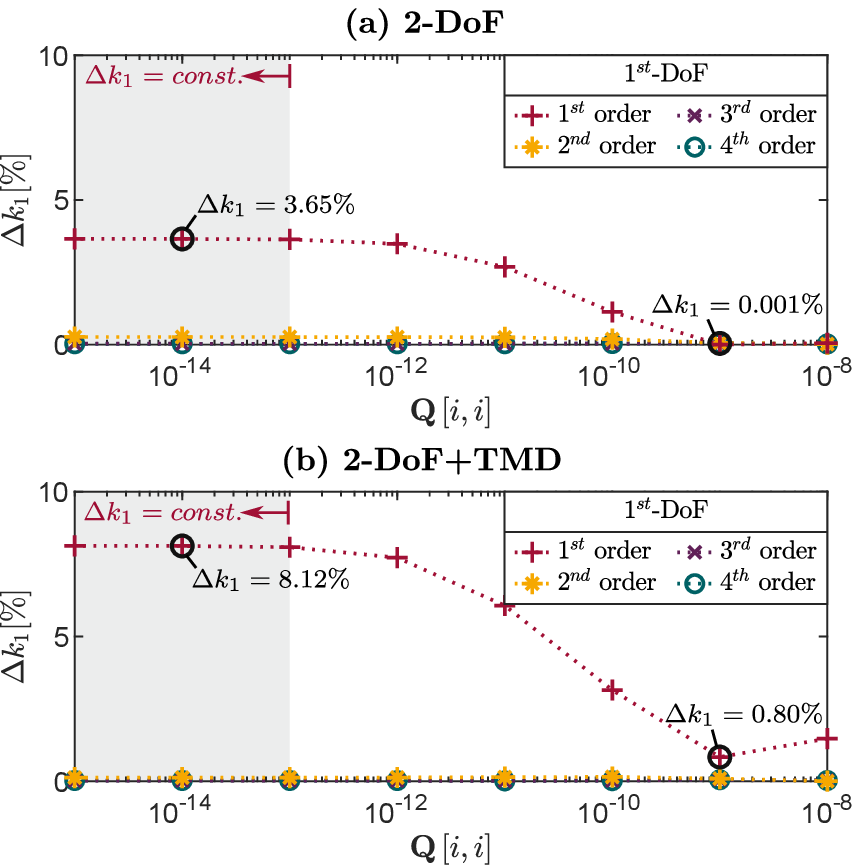}\\
	\caption{Study 3: Comparison of the stiffness deviations $\Delta k_1$ for $1^{st}$-$4^{th}$ order discretizations and variable system noise covariance levels $\mathbf{Q}\left\lbrack i,i\right\rbrack$. Estimations are calculated for the (a) 2-DoF and (b) 2-DoF+TMD systems under El Centro earthquake.}
	\label{Fig:12_Study_3}
\end{figure}
\begin{figure}[!htb]
	\centering
	\includegraphics[width=0.5\textwidth]{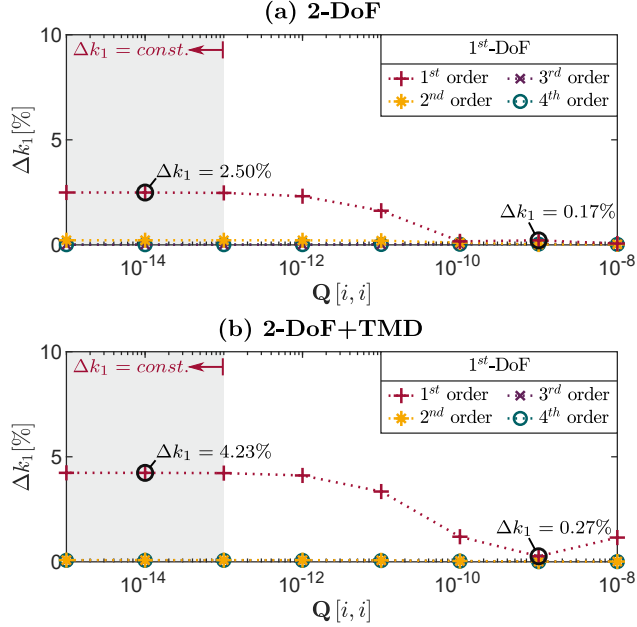}\\
	\caption{Study 3: Comparison of the stiffness deviations $\Delta k_1$ for $1^{st}$-$4^{th}$ order discretizations and variable system noise covariance levels $\mathbf{Q}\left\lbrack i,i\right\rbrack$. Estimations are calculated for the (a) 2-DoF and (b) 2-DoF+TMD systems under Northridge earthquake.}
	\label{Fig:13_Study_3}
\end{figure}

Figure \ref{Fig:11_Study_3} compares for the El Centro earthquake the estimated time histories of the $1^{st}$ DoF stiffness $\hat{k}_1$ with the true values $k_1$. Two different system noise covariance levels $\mathbf{Q}\left\lbrack i,i\right\rbrack$ are shown. At $\mathbf{Q}\left\lbrack i,i\right\rbrack=10^{-14}$ (left), the $1^{st}$ order Taylor expansion based model of the 2-DoF+TMD system causes larger deviations than the model of the 2-DoF without TMD. These results show the increased sensitivity of the system identification due to supplementary TMD. By increasing the covariance level to $\mathbf{Q}\left\lbrack i,i\right\rbrack=10^{-9}$ (right) the deviation reduces. In Figure \ref{Fig:11_Study_3}, the other investigated higher order models do not show any dependency with the covariance level.

The study is expended in Figure \ref{Fig:12_Study_3} for further $\mathbf{Q}\left\lbrack i,i\right\rbrack$ values. Here, we observe that the Taylor $1^{st}$ order expansion based model of the 2-DoF structure allows for system noise covariance level values higher than $\mathbf{Q}\left\lbrack i,i\right\rbrack=10^{-9}$ a high accuracy system identification with $\Delta k_1 <\SI{0.001}{\%}$. With the same order of the model, the system identification accuracy of the 2-DoF+TMD system also increases by increasing the system noise covariance. However, after reaching its minimum deviation at $\mathbf{Q}\left\lbrack i,i\right\rbrack=10^{-9}$ with increasing system noise covariance the deviation of the stiffness estimation increases again. This effect exists invisible small also for the 2-DoF structure without TMD.

Corresponding to the results of Figure \ref{Fig:11_Study_3}, also in Figure \ref{Fig:12_Study_3}, we see again for higher order models that the accuracy is independent from the system noise covariance level. Accordingly, as introduced before, we emphasize also with these results the necessity of higher order mathematical models for the identification MDoF+TMD systems.

In Figure \ref{Fig:13_Study_3}, the study is repeated for the near field Northridge earthquake. The performance results of the investigated models conform with the conclusions of the in Figure \ref{Fig:11_Study_3} and \ref{Fig:12_Study_3} shown El Centro results. Also here the course of the deviation parameter $\Delta k_1$ shows for the $1^{st}$ order Taylor expansion model of the 2-DoF structure a stable accuracy after a certain system noise level. On the other hand, for the same order 2-DoF+TMD model the deviation $\Delta k_1$ fluctuates depending on the system noise level. For the estimated $2^{nd}$ DoF stiffness $\hat{k}_2$ we get similar results, which we do not include here for the sake of brevity.


\section{Conclusions}\label{sec4}

In this paper, for MDoF structures with TMDs a recursive system identification method is presented, which is able to detect and localize abrupt stiffness changes during sudden events, such as earthquakes. The method enhances the UKF by a new adaptation formulation, which is modifying the state covariance initiated by a trigger parameter. The proposed adaptation algorithm operates in a recursive manner and calculates the trigger parameter depending on the innovation error, which is normalized by the measurement noise covariance. A constant threshold is formulated based on the sensors. Three parametric studies are conducted on a 2DoF+TMD system to investigate the performance of the system identification method. In the first study, earthquake, impulse and white noise excitations are applied. Single and combined abrupt stiffness changes of the DoFs of the structure are simulated. Time histories of estimated and true values of structural motion and stiffness changes are compared. Results show that the proposed identification method is able to detect and localize the abrupt stiffness changes. The estimated state conforms with the true values. The second study investigates the effects of the state covariance. On the 2DoF+TMD system an earthquake excitation is applied. An increase of the state covariance improves the parameter estimation performance. However, after a certain value, a further increase causes the identification method to behave oversensitive and loose its accuracy. The results conclude the necessity of an adaptive formulation of the state covariance as applied in the proposed approach. In the third study, the effects of the modeling accuracy are investigated on the 2DoF+TMD system under earthquake excitation. Besides the effects of the system noise covariance, the study considers also the effects of convergence orders for discretization using Taylor expansion. The results confirm that the identification of abrupt stiffness changes requires a high-level accuracy of the method, in particular, for the identification of MDoF structures with supplementary TMDs.



\bibliography{Ref}

\begin{thebibliography}{10}
\expandafter\ifx\csname url\endcsname\relax
  \def\url#1{\texttt{#1}}\fi
\expandafter\ifx\csname urlprefix\endcsname\relax\def\urlprefix{URL }\fi
\expandafter\ifx\csname href\endcsname\relax
  \def\href#1#2{#2} \def\path#1{#1}\fi

\bibitem{Devin.2019}
A.~Devin, P.~J. Fanning, Non-structural elements and the dynamic response of
  buildings: A review, Engineering Structures 187 (2019) 242--250.

\bibitem{Brincker.2015}
R.~Brincker, C.~E.~H. Ventura, Introduction to operational modal analysis, John
  Wiley {\&} Sons Inc, Chichester, West Sussex, 2015.

\bibitem{Soderstrom.1994}
T.~S{\"o}derstr{\"o}m, P.~Stoica, System identification, Prentice-Hall, New
  York, NY, 1994.

\bibitem{Brincker.2001}
R.~Brincker, L.~Zhang, P.~Andersen, Modal identification of output-only systems
  using frequency domain decomposition, Smart Materials and Structures 10~(3)
  (2001) 441--445.

\bibitem{Schleiter.2018}
S.~Schleiter, O.~Altay, S.~Klinkel, Experimental incremental system
  identification method using separate time windows on basis of ambient
  signals, in: J.~P. Conte, R.~Astroza, G.~Benzoni, G.~Feltrin, K.~J. Loh,
  B.~Moaveni (Eds.), Experimental vibration analysis for civil structures,
  Springer, 2018, pp. 694--704.

\bibitem{Smyth.1999}
A.~W. Smyth, S.~F. Masri, A.~G. Chassiakos, T.~K. Caughey, On-line parametric
  identification of mdof nonlinear hysteretic systems, Journal of Engineering
  Mechanics 125~(2) (1999) 133--142.

\bibitem{Chatzi.2009}
E.~N. Chatzi, A.~W. Smyth, The unscented kalman filter and particle filter
  methods for nonlinear structural system identification with non-collocated
  heterogeneous sensing, Structural Control and Health Monitoring 16~(1) (2009)
  99--123.

\bibitem{Kalman.1960}
R.~E. Kalman, A new approach to linear filtering and prediction problems,
  Journal of Basic Engineering 82~(1) (1960) 35.

\bibitem{Jazwinski.1997}
A.~H. Jazwinski, Stochastic processes and filtering theory, Vol.~64 of
  Mathematics in science and engineering, {Acad. Press}, San Diego, 1997.

\bibitem{Haykin.2001}
S.~S. Haykin, Kalman Filtering and Neural Networks, Wiley, New York, 2001.

\bibitem{Julier.1997}
S.~J. Julier, J.~K. Uhlmann, New extension of the kalman filter to nonlinear
  systems, in: I.~Kadar (Ed.), AeroSense '97, SPIE Proceedings, SPIE, 1997, p.
  182.

\bibitem{Julier.2004}
S.~J. Julier, J.~K. Uhlmann, Unscented filtering and nonlinear estimation,
  Proceedings of the IEEE 92~(3) (2004) 401--422.

\bibitem{Wan.14Oct.2000}
E.~A. Wan, R.~{van der Merwe}, The unscented kalman filter for nonlinear
  estimation, in: S.~S. Haykin (Ed.), The IEEE 2000 Adaptive Systems for Signal
  Processing, Communications, and Control Symposium., Piscataway, NJ, 2000, pp.
  153--158.

\bibitem{Hoshiya.1984}
M.~Hoshiya, E.~Saito, Structural identification by extended kalman filter,
  Journal of Engineering Mechanics 110~(12) (1984) 1757--1770.

\bibitem{Miah.2015}
M.~S. Miah, E.~N. Chatzi, F.~Weber, Semi-active control for vibration
  mitigation of structural systems incorporating uncertainties, Smart Materials
  and Structures 24~(5) (2015) 055016.

\bibitem{Roffel.2014}
A.~J. Roffel, S.~Narasimhan, Extended kalman filter for modal identification of
  structures equipped with a pendulum tuned mass damper, Journal of Sound and
  Vibration 333~(23) (2014) 6038--6056.

\bibitem{Miah.2017}
M.~S. Miah, E.~N. Chatzi, V.~K. Dertimanis, F.~Weber, Real-time experimental
  validation of a novel semi-active control scheme for vibration mitigation,
  Structural Control and Health Monitoring 24~(3) (2017) e1878.

\bibitem{Wu.2007}
M.~Wu, A.~W. Smyth, Application of the unscented kalman filter for real-time
  nonlinear structural system identification, Structural Control and Health
  Monitoring 14~(7) (2007) 971--990.

\bibitem{Erazo.2018}
K.~Erazo, S.~Nagarajaiah, Bayesian structural identification of a hysteretic
  negative stiffness earthquake protection system using unscented kalman
  filtering, Structural Control and Health Monitoring 25~(9) (2018) e2203.

\bibitem{Yang.2006}
J.~N. Yang, S.~Lin, H.~Huang, L.~Zhou, An adaptive extended kalman filter for
  structural damage identification, Structural Control and Health Monitoring
  13~(4) (2006) 849--867.

\bibitem{Lei.2016}
Y.~Lei, H.~Zhou, Z.-L. Lai, A computationally efficient algorithm for real-time
  tracking the abrupt stiffness degradations of structural elements,
  Computer-Aided Civil and Infrastructure Engineering 31~(6) (2016) 465--480.

\bibitem{Bisht.2014}
S.~S. Bisht, M.~P. Singh, An adaptive unscented kalman filter for tracking
  sudden stiffness changes, Mechanical Systems and Signal Processing 49~(1-2)
  (2014) 181--195.

\bibitem{Rahimi.2017}
A.~Rahimi, K.~D. Kumar, H.~Alighanbari, Fault estimation of satellite reaction
  wheels using covariance based adaptive unscented kalman filter, Acta
  Astronautica 134 (2017) 159--169.

\bibitem{BarShalom.2001}
Y.~Bar-Shalom, X.-R. Li, T.~Kirubarajan, Estimation with applications to
  tracking and navigation, New York, 2001.

\bibitem{Bendat.2010}
J.~S. Bendat, A.~G. Piersol, Random data: Analysis and measurement procedures,
  fourth edition Edition, {John Wiley {\&} Sons, Inc}, Hoboken, New Jersey,
  2010.

\bibitem{Warburton.1982}
G.~B. Warburton, Optimum absorber parameters for various combinations of
  response and excitation parameters, Earthquake Engineering {\&} Structural
  Dynamics 10~(3) (1982) 381--401.

\end{thebibliography}

\end{document}